\documentclass[11pt]{article}
\def\half{\frac{1}{2}}

\newfont{\bbbold}{msbm10 scaled \magstep1}

\def\cA{{\cal A}}

\def\cD{{\cal D}}
\def\cE{{\cal E}}
\def\cF{{\cal F}}

\def\cR{\cal R}

\def\cW{\cal W}

\newfont{\goth}{eufm10 scaled \magstep1}

\def\a{\alpha}
\def\b{\beta}
\def\c{\gamma}
\def\d{\delta}
\def\e{\epsilon}
\def\f{\phi}\def\F{\Phi}
\def\h{\eta}
\def\i{\iota}
\def\k{\kappa}
\def\L{\Lambda}
\def\m{\mu}
\def\n{\nu}

\def\r{\rho}
\def\s{\sigma}\def\S{\Sigma}
\def\t{\tau}
\def\th{\theta}

\def\beq{\begin{equation}}\def\eeq{\end{equation}}
\def\beqa{\begin{eqnarray}}\def\eeqa{\end{eqnarray}}
\def\barr{\begin{array}}\def\earr{\end{array}}

\def\O{\Omega}
\def\del{\partial}
\def\ua{\underline{\alpha}}
\def\ub{\underline{\phantom{\alpha}}\!\!\!\beta}

\def\unth{\underline{\theta}}

\def\una{\underline a}\def\unA{\underline A}
\def\unb{\underline b}\def\unB{\underline B}
\def\unc{\underline c}\def\unC{\underline C}

\def\unM{\underline M}
\def\unN{\underline N}

\def\unx{\underline{x}}  \def\unz{\underline{z}}

\def\xz{\times}

\def\nab{\nabla}

\def\hE{\hat{E}}
\def\nno{\nonumber}

\def\hb{\widehat{b}}
\def\hf{\widehat{f}}
\def\hv{\widehat{v}}\def\hz{\widehat{z}}
\def\hA{\widehat{A}}\def\hB{\widehat{B}}\def\hC{\widehat{C}}\def\hE{\widehat{E}}
\def\hH{\widehat{H}}
\def\hM{\widehat{M}}\def\hN{\widehat{N}}

\def\hW{\widehat{W}}

\def\gha{\widehat{\a}}\def\ghb{\widehat{\b}}\def\ghc{\widehat{\c}}\def\ghd{\widehat{\d}}

\def\ghm{\widehat{\m}}\def\ghn{\widehat{\n}}\def\ghr{\widehat{\r}}\def\ghs{\widehat{\s}}

\def\tb{\widetilde{b}}
\def\tB{\widetilde{B}}

\let\la=\label

\def\nn{\nonumber}
\def\bd{\begin{document}}
\def\ed{\end{document}}
\def\ba{\begin{array}}
\def\ea{\end{array}}
\def\bea{\begin{eqnarray}}
\def\eea{\end{eqnarray}}
\def\ft#1#2{{\textstyle{{\scriptstyle #1}\over {\scriptstyle #2}}}}
\def\fft#1#2{{#1 \over #2}}
\newcommand{\be}{\begin{equation}}
\newcommand{\ee}{\end{equation}}
\newcommand{\eq}[1]{(\ref{#1})}
\newcommand{\w}[1]{\\[0.#1cm]}
\def\eqs#1#2{(\ref{#1}-\ref{#2})}
\def\det{{\rm det\,}}
\def\tr{{\rm tr}}
\newcommand{\hoch}[1]{$\, ^{#1}$}
\newcommand{\tamphys}{\it\small Center for Theoretical Physics,
Texas A\&M University, College Station, TX 77843, USA}
\newcommand{\kings}
{\it\small Department of Mathematics, King's College, London, UK}
\newcommand{\uu}
{\it\small Department of Theoretical Physics, Uppsala, Sweden}
\newcommand{\hip}
{\it\small HIP-Helsinki Institute of Physics,
P.O. Box 64 FIN-00014 University of Helsinki, Suomi-Finland}
\newcommand{\stock}
{\it\small Department of Theoretical Physics, Stockholm, Sweden}
\makeatletter
\renewcommand\theequation{\thesection.\arabic{equation}}
\@addtoreset{equation}{section} \makeatother

\newcommand{\auth}
{\large P.S. Howe\hoch{1,2}, U. Lindstr\"om\hoch{2,3} and L.
Wulff\hoch{4}}

\begin{document}

\hfill{KCL-TH-05-03}

\hfill{UUITP-08/05}

\hfill{HIP-2005-17/TH}

\hfill{USITP-05-01}

\hfill{hep-th/0505067}

\hfill{\today}

\vspace{20pt}

\begin{center}
{\Large{\bf Superstrings with boundary fermions}}
\vspace{30pt}

\auth

\vspace{15pt}

\begin{itemize}
\item [$^1$] \kings \item [$^2$] \uu \item  [$^3$] \hip \item  [$^4$] \stock
\end{itemize}

\vspace{60pt}

{\bf Abstract}

\end{center}

The Green-Schwarz action for an open superstring with additional boundary
fermions, representing Chan-Paton factors, is studied at the classical level.
The boundary geometry is described by a bundle $\widehat M$, with fermionic
fibres, over the super worldvolume $M$ of a D-brane  together with a map from
$\widehat M$ into the $N=2, D=10$ target superspace $\unM$. This geometry is
constrained by the requirement of kappa-symmetry on the boundary together with
the use of the equations of motion for the fermions. There are two constraints
which are formally similar to those that arise in the abelian case but
which differ because of the dependence on the additional coordinates. The
model, when quantised, would be a candidate for a fully kappa-symmetric
theory of a stack of coincident D-branes including a non-abelian Born-Infeld
sector. The example of the D9-brane in a flat background is studied. The
constraints on the non-abelian field strength are shown to be in agreement with
those derived from the pure spinor approach to the superstring. A covariant
formalism is developed and the problem of quantisation is discussed.

\pagebreak \tableofcontents \setcounter{page}{1}


\section{Introduction}


Born-Infeld dynamics has turned out to play a major r\^ole in string theory.
This was first observed in studies of the open string sigma model
\cite{Fradkin:1985qd,Abouelsaood:1986gd,Bergshoeff:1987at,Leigh:1989jq}
but has come to
greater prominence in the context of D-branes. As is well-known, the effective
Lagrangian for a single D-brane in type II string theory consists of a
supersymmetric Dirac-Born-Infeld kinetic term together with an appropriate
Wess-Zumino term which involves the RR fields and the worldvolume
electromagnetic field strength tensor suitably modifed by the NS two-form. An
intriguing problem which is still not fully understood is what the non-abelian
version of this theory is. This theory should describe the effective
dynamics of a set of coincident D-branes. In this paper we shall outline an
approach to this problem based on demanding kappa-symmetry  for a Green-Schwarz
superstring action with boundary fermions which represent Chan-Paton factors.

Many features of the bosonic terms in the action are known. In the simplest
generalisation of the Born-Infeld action to the non-abelian case, proposed in
\cite{Tseytlin:1997cs}, the electromagnetic field strength tensor is replaced
by a non-abelian one
and the Yang-Mills indices are dealt with by taking the symmetrised trace in
each term in the power series expansion of the Lagrangian. It is known from
analyses of string scattering amplitudes \cite{Kitazawa:1987xj}
that there are
other terms in the effective action of the open string, but the Tseytlin action
is expected to provide at least a part of the non-abelian Born-Infeld theory.
Dimensional reduction of this action together with
T-duality yields a lot of information about terms in the action involving scalar
fields and also about the Wess-Zumino term \cite{Myers:1999ps}. In
particular, one sees the emergence of non-commutativity (the transverse scalars
become matrix valued) and the interesting feature that lower-dimensional branes
can detect higher-dimensional RR fields. These features were also
found from studies of D-branes using matrix models
\cite{Taylor:1999gq,Taylor:1999pr}.

\pagebreak

The results referred to above do not include any fermion fields or make any
reference to supersymmetry. Before discussing this point further in the
non-abelian case it is perhaps worth pointing out that the effective action for
a single D-brane in the Born-Infeld approximation is completely fixed by
supersymmetry. This can perhaps be seen most elegantly in the superembedding
formalism \cite{Sorokin:1988nj,Bandos:1995zw,Sorokin:1999jx}
where a single constraint on the embedding of the super worldvolume of
the brane into the super target space determines the structure of the
worldvolume multiplet, the dynamics of this multiplet and implies that the
background should be an on-shell supergravity solution
\cite{Howe:1996mx,Howe:1997wf}. In cases of low codimension one must also
impose a constraint on the modified field strength tensor $\cF$
\cite{Howe:2000vk,Drummond:2001uj}
which can be
derived from the superembedding constraint in the other cases. Both constraints
can be derived from an analysis of superstrings ending on branes, either in a
hybrid GS approach \cite{Chu:1998jv}, or completely within the superembedding
formalism \cite{Chu:1998pb}.

There have been various attempts to generalise the above to the non-abelian
case. A  model with matrix-valued kappa-symmetry was proposed in
\cite{Bergshoeff:2000kt}, although it
does not seem to be quite right \cite{Bergshoeff:2001dc}.
Actions with a single
kappa-symmetry incorporating non-abelian multiplets
interacting with the geometry of a brane have been constructed in lower
dimensional target superspaces
\cite{Sorokin:2001av,Sorokin:2002je,Drummond:2002kg}.
There have also been studies of
supersymmetric non-abelian Born-Infeld  theory from a purely field-theoretic
point of view \cite{Gonorazky:1998ay,Ketov:2000fv,Refolli:2002gt}.
The supersymmetrisation of a number of
terms known to be present in the effective action of open strings was discussed
in components in \cite{deRoo:2002ap,deRoo:2003xv} and in $N=4$ superfields in
\cite{Drummond:2003ex}.
In particular, these studies made contact with the proposal for the purely
bosonic sector based on T-duality put forward in
\cite{Koerber:2002zb}, and with results derived from scattering amplitude
calculations \cite{Medina:2002nk,Chandia:2003sh,Barreiro:2005hv}.
Another possible approach to
non-abelian Born-Infeld dynamics is via the study of boundary couplings in the
RNS sigma model \cite{Hassan:2000zk,Hassan:2003uq}.

The subject has also been discussed in the pure
spinor approach to string theory. In particular, Berkovits and Pershin derived
some terms involving non-abelian fields in the background of a single brane
using boundary fermions to represent Chan-Paton factors and working at the
classical level with respect to these fields \cite{Berkovits:2002ag}
(see also \cite{Schiappa:2005mk}).
The idea of using boundary fermions in this way was introduced by \cite{Marcus:1986cm}
and studied
further in \cite{Dorn:1996an,Kraus:2000nj}.
In the current paper we shall adopt the philosophy of Berkovits
and Pershin but in the context of the Green-Schwarz string. We generalise in
several ways. We discuss branes of arbitrary dimension, we allow for arbitrary
supergravity backgrounds, we work to all orders in the boundary fermions and we
use covariant methods in as far as this is possible. This approach therefore
defines a particular approximation to the effective string action with
non-abelian fields. In principle, we could obtain a better
approximation by quantising the fermions. One might propose the result of this
quantisation as a definition of what one means by non-abelian Born-Infeld
dynamics for a set of coincident branes. As is well-known, it is not
self-evident what this approximation should be because some derivative terms can
be replaced with commutators in the non-abelian theory.

As we shall briefly discuss at the end of the paper, it is not entirely
straightforward to carry out the quantisation of the fermions by themselves. The
main reason for this is that the geometrical picture underlying the construction
is slightly unusual. In the case of a single D-brane one takes the
(Green-Schwarz) sigma model field to map into the super worldvolume of the brane
on the boundary of the string. In the present case, following Berkovits and
Pershin, we shall take the boundary sigma model field to depend on the boundary
fermions as well. One way of thinking about this would be to replace the
type II target space by a space which has additional fermionic fibres which are
localised on the brane. This is a somewhat singular construction but one could
view it as the limit of a thickened brane. This would be reasonable on physical
grounds since there are supposed to be several branes which coincide. Moreover,
the supergravity brane solution also has a natural thickness.  However, from the
point of view of quantising the model, this picture implies that there is a
hidden dependence on the fermions residing in the bulk part of the action.
Another way of thinking about this geometry is that it roughly corresponds to
the notion of a coherent sheaf, a topic that has been discussed in the boundary
fermion picture in the literature \cite{Govindarajan:2001kr}.

We shall not address the problem of quantisation to any great extent in this
paper, postponing a fuller discussion to a later date. However, there are
several nice features that can be observed in the classical boundary fermion
approximation. In particular, we can keep control of all of the symmetries of
the problem in a systematic fashion. These include the one-form gauge symmetry
associated with the NS $B$-field for which we can derive the appropriate
modification of the non-abelian field strength tensor which respects this
symmetry. We also discuss the diffeomorphisms of the super worldvolume $M$.
Since this is now the base of a larger space, $\hM$, which includes the
additional fermions as fibre coordinates, these diffeomorphisms naturally depend
on all of the coordinates. This is required in order for us to be able to gauge
away all of the non-physical degrees of freedom and is related to the idea
of matrix-valued kappa-symmetry introduced in \cite{Bergshoeff:2000kt}. These
features are all derived systematically from the action in the next section. It
is also shown that the requirement of kappa-symmetry on the boundary of the
string  leads to geometrical conditions on what one might call the generalised
superembedding defined by the map from $\hM$ into the super target space $\unM$.
These constraints are remarkably similar to those for a single brane, although
it is somewhat more difficult to work out the consequences. In section four we
do this for the D9-brane  in a flat target space and relate our results
explicitly to those of Berkovits and Pershin. In section five we propose a
completely covariant formulation of the generalised superembedding. This version
is covariant with respect to diffeomorphisms of $\hM$. In section six we
briefly discuss the quantisation problem and give our conclusions.


\section{Action with boundary fermions}


\subsection{Action and variation}

The Green-Schwarz action for the superstring is described by a map $f_o$ from a
bosonic worldsheet $\S$ to an $N=2, D=10$ target superspace $\unM$. We shall
suppose that the boundary of the worldsheet, $\del\S$, maps to a space, $\hM$,
which is a bundle over the super worldvolume, $M$, of a D-brane, the
fibres of this bundle being fermionic. There is a map, $\hf:\hM \rightarrow
\unM$ which extends a superembedding, $f:M\rightarrow \unM$, of the brane into
the target superspace. The interpretation of this set-up is that the
brane is actually the centre-of-mass of a set of coincident D-branes while the
fermionic coordinates represent Chan-Paton factors. As we shall see, this allows
us to introduce non-abelian gauge fields on the brane.

The worldsheet coordinates are denoted by $y^i,i=0,1$ and the coordinate on the
boundary is denoted by $t$. The coordinates of the superspace $M$ are
$z^M=(x^m,\th^\m)$ and the coordinate basis one-forms $dz^M$  are related to
preferred basis one-forms $E^A=(E^a,E^\a)$ by means of the supervielbein,
$E^A=dz^M E_M{}^A$. Similar conventions are used for $\unM$, with the difference
that all indices are underlined. Primed indices $A'=(a',\a')$ denote
directions normal to the brane in $\unM$. The coordinates for $\hM$ are denoted
by $z^{\hM}=(z^M,\h^{\ghm})$, where the index $\ghm$, which labels the boundary
fermions, can take on $n$ values. When coordinates are used as arguments of
functions we shall write $z,\hz,\unz$  for functions on $M,\hM,\unM$
respectively.

The bulk part of the GS action is the same as usual, but we shall allow for
a general supergravity background which would in any case be required by
kappa-symmetry. In our conventions this action is

\begin{equation}
S_{\rm bulk}=-\int_{\S}\, d^2\,y\,\left(\sqrt{-g} +\half \e^{ij} B_{ij}\right)
\ ,
\la{2.1}
\end{equation}

where $g_{ij}$ is the induced GS metric on $\S$,

\begin{equation}
g_{ij}:=E_i{}^{\una}E_j{}^{\unb} \h_{\una\unb}\ .
\la{2.2}
\end{equation}

The notation $E_i{}^{\unA}$ denotes the pull-back map from $\unM$ to $\S$ in a
preferred basis,

\begin{equation}
E_i{}^{\unA}:=\del_i z^{\unM} E_{\unM}{}^{\unA}\ .
\la{2.3}
\end{equation}

Similarly, $B_{ij}$ denotes the pull-back of the NS two-form potential,

\begin{equation}
B_{ij}:= \del_j z^{\unN}\del_i z^{\unM} B_{\unM\unN}=  E_j{}^{\unB}E_i{}^{\unA}
B_{\unA\unB}\ .
\la{2.4}
\end{equation}

The variation of the bulk action gives

\bea
\d S_{\rm bulk} &=&\int_{\S}\,d^2 y\,\left\{\sqrt{-g}g^{ij}\left (\d z^{\una}
\nab_i E_{j\una}+\d z^{\unA} E_i{}^{\unB} T_{\unB\unA}{}^{\unc} E_{j\unc}\right
)+\half\d z^{\unA}\e^{ij}E_i{}^{\unB} E_{j}{}^{\unC} H_{\unC\unB\unA}
\right\}\nno\w2
 &+& \int_{\S}\,d^2y\, \del_i\left (- \sqrt{-g}g^{ij}\d z^{\una}
E_{j\una} +\e^{ij} \d z^{\unA} E_j{}^{\unB} B_{\unB\unA}\right )
\la{2.5}
\eea

In this equation $\d z^{\unA}:=\d z^{\unM} E_{\unM}{}^{\unA}$ denotes the
variation referred to a preferred basis, $\nab_i$ is a covariant derivative with
respect to both target space and worldsheet indices, $T_{\unA\unB}{}^{\unC}$ is
the target space torsion and $H=dB$ is the NS three-form field strength.
Bosonic target space indices can be raised and lowered by means of
$\h_{\una\unb}$. In the rest of the paper we shall only be concerned with the
second line of the right-hand-side of this equation which will contribute to the
boundary theory. This contribution is

\begin{equation}
\d S_{\rm bulk-bdry}=\int\, dt\,n_i \left(-g^{ij}\d z^{\una}
E_{j\una} +\frac{\e^{ij}}{\sqrt{-g}} \d z^{\unA} E_j{}^{\unB} B_{\unB\unA}\right
) \la{2.6}
\end{equation}

where $n_i$ denotes a unit normal to $\del\S$. This can be rewritten as

\begin{equation}
\d S_{\rm bulk-bdry}=\int\, dt\, \left(-\d z^{\unM} E_{\unM}{}^{\una}
E_{1\una} +\d z^{\unM} \dot z^{\unN} B_{\unN\unM}\right )
\la{2.7}\ .
\end{equation}

where

\begin{equation}
E_{1\una}:=n_i g^{ij} E_{j\una}\ .
\la{2.7.1}
\end{equation}

In the case of a single D-brane, the mapping would be restricted to the
worldvolume of the brane on the boundary. In this case, however, we shall assume
that the boundary field takes its values in $\hM$. Thus, on the boundary,
$z^{\unM}= z^{\unM}(\hz)=z^{\unM}(z^M,\h^{\ghm})$. Taking this into account, we 
find that \eq{2.7} can be written

\begin{equation}
\d S_{\rm bulk-bdry}=\int\, dt\, \d z^{\hM}\del_{\widehat{M}}z^{\unM} \left(-
E_{\unM}{}^{\una} E_{1\una} + \dot z^{\hN}\del_{\hN} z^{\unN} B_{\unN\unM}\right
) \la{2.8}\ .
\end{equation}

The action for the boundary fermions is

\begin{equation}
S_{\rm bdry}=\int_{\del\S}\, \cA
\la{2.9}
\end{equation}

where $\cA$ is the pull-back of a one-form potential on $\hM$. The variation of
this action gives

\begin{equation}
\d S_{\rm bdry}=-\int_{\del\S}\,dt\,\d z^{\hM} \dot z^{\hN} (d\cA)_{\hN\hM}\ .
\la{2.10}
\end{equation}

The
combined action, $S_{\rm tot}:=S_{\rm bulk}+S_{\rm bdry}$, has a lot of
symmetries. In particular, it is covariant under the local geometrical
symmetries of the target space, including diffeomorphisms of $\hM$, and the
latter depend on all of the coordinates including the Chan-Paton
fermions. There is a $U(1)$ gauge symmetry, $\cA\mapsto \cA + d a$, as well as
a manifest abelian one-form gauge symmetry, $B\mapsto B + db$, $\cA\mapsto \cA
+b$, where appropriate pull-backs are understood. The non-abelian gauge
symmetry will emerge later from the $U(1)$ and the vertical diffeomorphisms of
$\hM$.

The equation of motion for the boundary fermions, following from \eq{2.8} and
\eq{2.10},  is

\begin{equation}
\dot \h^{\ghm}=-\left(\dot z^N K_{N\ghn} + \del_{\ghn} z^{\unM}
E_{\unM}{}^{\una} E_{1\una}\right)N^{\ghm\ghn}\ ,
\la{2.11}
\end{equation}

where

\be
K:=d\cA - \hf^*B
\la{2.11.1}
\ee

and $N^{\ghm\ghn}$ is the inverse of
$K_{\ghm\ghn}$. Using this equation in the total boundary variation (again from
\eq{2.8} and \eq{2.10}) we find that the residual variation is

\begin{equation}
\d_{\rm bdry} S_{\rm tot}=-\int_{\del\S}\, dt\, \d z^N\left\{\dot z^M
\left(K_{MN}-K_{M\ghm}N^{\ghm\ghn} K_{\ghn N}\right) +
\cD_N z^{\unM} E_{\unM}{}^{\una} E_{1\una}\right\}\ ,
\la{2.12}
\end{equation}

where we have introduced a covariant derivative (vector field)

\begin{equation}
\cD_M:= \del_M- K_M{}^{\ghm}\del_{\ghm};\qquad {\rm where}\qquad
K_M{}^{\ghm}:=K_{M\ghn}N^{\ghn\ghm}\ .
\la{2.13}
\end{equation}

The object in the round brackets on the right-hand-side of \eq{2.12} will play
a central r\^ole in the following and so we give it a new name, $\cF_{MN}$,

\begin{equation}
\cF_{MN}:= K_{MN}-K_{M\ghm}N^{\ghm\ghn} K_{\ghn N}\ .
\la{2.14}
\end{equation}


\subsection{Kappa-symmetry}


The total action will be required to be invariant under kappa-symmetry. This
can be discussed for the bulk and boundary separately. Kappa-symmetry  in the
bulk is of course well-understood and is ensured by requiring the background
superspace to be an arbitrary supergravity background, i.e. the background
fields are those of the supergravity multiplet and they are on-shell. It has
been discussed for type II strings in refs \cite{Grisaru:1985fv,Bergshoeff:1997cf}.
It is the 
boundary kappa-symmetry which is of most interest in this paper. We shall define
a boundary kappa-symmetry transformation to be (the leading component of) an odd
diffeomorphism of the worldvolume of the brane, $M$. This is the
usual superembedding interpretation of kappa-symmetry. In order to do this we
have to select an odd sub-bundle $F$ of the tangent bundle $TM$, and to do this
we have to make use of a worldvolume supervielbein $E_M{}^A$. We then set

\be
\d_{\k}z^{\a}=v^{\a}\qquad \d_{\k} z^{a}=0 \ ,
\la{2.15}
\ee

where $\d z^A:=\d z^M E_M{}^A$. The odd diffeomorphism parameter $v^{\a}$ can be
related to the kappa-symmetry parameter in the usual fashion,

\begin{equation}
\k^{\ua}:=v^{\a} E_{\a}{}^M \del_M z^{\unM} E_{\unM}{}^{\ua}
\la{2.16}
\end{equation}

If we substitute this variation into \eq{2.12} we see that kappa-symmetry on the
boundary will be assured provided that two geometrical conditions are
satisfied,

\be
\cE_{\a}{}^{\una}=0
\la{2.17}
\ee
and
\be
\cF_{\a B}=0
\la{2.18}
\ee

where

\begin{equation}
\cE_A{}^{\unA}:= E_A{}^M \cD_M z^{\unM} E_{\unM}{}^{\unA}\ .
\la{2.19}
\end{equation}

The equations \eq{2.17} and \eq{2.18} are remarkably similar to the standard
equations describing single D-branes in the superembedding formalism. They were
derived in a GS approach in \cite{Chu:1998jv}
and can also be understood in terms of
superembeddings of branes ending on branes \cite{Chu:1998pb}. The first equation, the
superembedding constraint, is now generalised to include the covariant
derivative $\cD_M$, while the constraint on the two-form will contain
information about the non-abelian field strength tensor. As we shall see, these
two equations determine the structure of the multiplets on the brane and their
equations of motion.

It might seem that the form of the kappa-symmetry constraints depends on the
choice of odd tangent bundle on $M$. This is indeed the case, since if we made a
different choice, say $E'_{\a}=E_{\a} + \L_{\a}{}^a E_{a}$, where $\L$ is some
field, then the form of the constraints would change. A better statement
would be to say that kappa-symmetry implies that there exists a choice of odd
tangent bundle $F$ on the brane such that equations \eq{2.17} and \eq{2.18}
hold.


\section{Geometrical interpretation}


In this section we shall give a geometrical interpretation of the results
derived above. The first observation is that the vector field $\cD_M$
\eq{2.13} defines horizontal subspaces in the tangent spaces of $\hM$. Thus
we see that the use of the equations of motion
for the fermions induces a bundle structure in $\hM$ with a connection defined
by the vector field $\cD_M$. It should be borne in mind, however, that this
structure is not quite the same as the Yang-Mills structure which will be
associated with a different covariant derivative (denoted by
$D_M$) and which will be discussed later. Moreover, as we shall see, the
horizontal subspaces are not preserved by $\h$-dependent diffeomorphisms of
$M$, so that the bundle structure is in some sense generalised. This
generalisation can be thought of as a consequence of maintaining general
covariance.

The field $\cF_{MN}$ \eq{2.14} is a
horizontal two-form which is in fact the horizontal projection of $K$. The
definition of the horizontal subspace makes use of $K$ and it can easily be
checked that, in the horizontal lift basis $(\cD_M,\del_{\ghm})$, $K$ has either
purely horizontal or purely vertical components. The dual basis is

\begin{equation}
dz^M \qquad {\rm and} \qquad e^{\ghm}:= d\h^{\ghm}+dz^M K_M{}^{\ghm}\ ,
\la{3.1}
\end{equation}

so that

\begin{equation}
K=\half \left( dz^N dz^M \cF_{MN} + e^{\ghn}e^{\ghm} K_{\ghm\ghn}\right)\ .
\la{3.2}
\end{equation}

The specification of a horizontal subspace allows us to define the notion of a
covariant pull-back of forms from $\unM$. This is just the horizontal part of
the ordinary pull-back. We shall denote such pull-backs by hats. Later on we
shall have a different covariant pull-back which will be denoted by a tilde. For
example, given a one-form $b$ on $\unM$ its covariant pull-back is

\begin{equation}
\hb_M := \cD_M z^{\unM} b_{\unM}\ .
\la{3.3}
\end{equation}

Higher-rank forms will, of course, have mixed components. For the $B$-field, for
example, as well as $\hB_{MN}$, we have

\begin{equation}
\hB_{M\ghn}:=\del_{\ghn}z^{\unN}  \cD_M z^{\unM} B_{\unM\unN}\ .
\la{3.4}
\end{equation}

We shall denote the purely horizontal pull-back of a form by $\hf$ in contrast
to the full pull-back which will be denoted by $\hf^*$.

The field $\cF_{MN}$ obeys a nice Bianchi identity which follows as a
consequence of its definition as the horizontal part of $K$, the fact that $K$
has no mixed component, and the Bianchi identity for $K$ which is

\begin{equation}
dK =-\hf^* H\ .
\la{3.5}
\end{equation}

The $\cF$ Bianchi is

\begin{equation}
\cD\cF=-\hf H\ ,
\la{3.6}
\end{equation}

or, in components,

\begin{equation}
3\cD_{[M} \cF_{NP]}=-\hH_{MNP}\ .
\la{3.7}
\end{equation}

It is instructive to check the consistency of \eq{3.7} directly. This makes use
of the curvature of $\cD_M$ which is defined by

\begin{equation}
[\cD_M,\cD_N]:=-{\cR}_{MN}{}^{\,\ghr}\,\del_{\ghr}=
N^{\ghr\ghs}\left(\del_{\ghs}{\cF}_{MN} + \hH_{\ghs MN}\right)\del_{\ghr}\
. \la{3.8}
\end{equation}

To show the consistency of \eq{3.7} one applies a second covariant derivative to
both sides, antisymmetrises and makes use of the definition of the curvature
$\cR$.

Now a choice of a horizontal subbundle in $T\hM$
is normally only preserved by diffeomorphisms of the base $M$ which do not
depend on the fibre coordinates, but in the present case we will shortly see
that the geometrical constraints which follow from kappa-symmetry are covariant
under $\h$-dependent diffeomorphisms of $M$. Let $v$ be a vector field on $\hM$
which generates an infinitesimal diffeomorphism. It can be written

\bea
v&=& v^M \del_M + v^{\ghm} \del_{\ghm}\nn\\
&=&  \hv^M \cD_M + \hv^{\ghm} \del_{\ghm}
\la{3.9}
\eea

from which

\bea
\hv^M&=& v^M \nn\\
\hv^{\ghm}&=&v^{\ghm} + v^M K_M{}^{\ghm}\ .
\la{3.10}
\eea

A short computation shows that

\bea
\d K_{M}{}^{\ghm}&=&\hv^N\left(\cD_N K_M{}^{\ghm}  -\cD_M
K_N{}^{\ghm}\right)-N^{\ghm\ghn}\del_{\ghn}\hv^N\cF_{NM} \nn\w2
&\phantom{=}& +\, \cD_M \hv^{\ghm}+\hv^{\ghn}\del_{\ghn}K_M{}^{\ghm}\ .
\la{3.11}
\eea

The transformation of a horizontal one-form is then found to be

\bea
\d \hb_M&=& \hv^N \cD_N \hb_M + \cD_M \hv^N \hb_N +
N^{\ghm\ghn}\del_{\ghn}\hv^N\cF_{NM}b_{\ghm}\nn\w2
&\phantom{=}& +\,\hv^{\ghm}\del_{\ghm}\hb_M\ .
\la{3.12}
\eea

The last term on the first line of the right-hand-side of this equation shows
that the violation of horizontality under such transformations is caused by the
$\h$-dependence of the $\hv^M$ parameter. However, as we remarked above, the
geometrical constraints implied by kappa-symmetry are covariant. It is
straightforward to show that $\cF$ transforms in a nice tensorial fashion. This
follows from the facts that $\cF$ is the horizontal part of $K$ and that the
latter has no mixed component in the horizontal lift basis. Thus the
undesired term involving $N$ which appears in \eq{3.12} does not appear in the
transformation of $\cF$, which is

\bea
\d\cF_{MN}&=& \hv^P \cD_P \cF_{MN}  +\cD_M \hv^P  \cF_{PN}   -
\cD_N \hv^P\cF_{PM}\nn\w2
&\phantom{=}&+\,\hv^{\ghm}\del_{\ghm} \cF_{MN}\ .
\la{3.13}
\eea

On the other hand, the
generalised superembedding condition  \eq{2.17} involves the horizontal
pull-back of the supervielbein, as can be seen from  equation \eq{2.19}. There
will therefore be a contribution to the transformation of $\cE_A{}^{\unA}$ of
the form

\begin{equation}
\d \cE_A{}^{\unA}\sim E_A{}^M
N^{\ghm\ghn}\del_{\ghn}\hv^N\cF_{NM}\del_{\ghm}z^{\unM} E_{\unM}{}^{\unA}\ .
\la{3.14}
\end{equation}

This looks slightly unpleasant, but under closer inspection we observe that
the superembedding condition \eq{2.17} transforms into the
$\cF$ constraint \eq{2.18}, while the latter transforms into itself. This
argument makes it plausible that the kappa-symmetry constraints are indeed
covariant under diffeomorphisms of $\hM$, but it is not complete as we have
not taken into account the transformation properties of the supervielbein. We
shall return to this point in the section on the covariant formalism.

We conclude this section with a discussion of  gauge-fixing and the
non-abelian gauge field. The transformation of the gauge field under $U(1)$
gauge transformations and diffeomorphisms can be written (in a coordinate basis)

\bea
\d \cA_M&=& \del_M a + v^N \left(\del_N \cA_M  -\del_M \cA_N\right)
+ v^{\ghn} \left(\del_{\ghn} \cA_M  -\del_M \cA_{\ghn}\right)\nn \w2
\d \cA_{\ghm}&=& \del_{\ghm} a + v^N \left(\del_N \cA_{\ghm}  -\del_{\ghm}
\cA_N\right) + v^{\ghn} \left(\del_{\ghn} \cA_{\ghm}  +\del_{\ghm}
\cA_{\ghn}\right)\ ,
\la{3.15}
\eea

where we have shifted the original $U(1)$ parameter $a$ by $(v^M \cA_M +
v^{\ghm} \cA_{\ghm})$. Provided that $\del_{\ghn} \cA_{\ghm}  +\del_{\ghm}
\cA_{\ghn}$ is non-singular we can use $v^{\ghm}$ to bring $\cA_{\ghm}$ to a
standard form which we shall take to be

\be
\cA_{\ghm}=\half \h_{\ghm}:= \half \d_{\ghm\ghn}\h^{\ghn}
\la{3.16}
\ee

In this gauge  residual vertical diffeomorphisms are given by

\be
v^{\ghm}=\d^{\ghm\ghn}\left(-\del_{\ghn} a + v^M A_{M\ghn}\right)
\la{3.17}
\ee

where

\be
A_{M\ghn}:= \del_{\ghn} \cA_M \ .
\la{3.18}
\ee

We shall now define $A_M$ to be $\cA_M$ in this gauge. This is the non-abelian
gauge field whose field strength tensor is

\be
F_{MN}:=\del_M A_N -  \del_N A_M + (A_M,A_N)\ ,
\la{3.19}
\ee

where the bracket $(,)$ can be thought of as a fibre Poisson bracket and is
defined by

\be
(f,g):= \d^{\ghm\ghn} \del_{\ghm} f \del_{\ghn} g\ .
\la{3.20}
\ee

The gauge transformation of $A_M$ is

\be
\d A_M=\del_M a + (A_M,a)\ .
\la{3.21}
\ee

The gauge group is thus the symplectic group of the fibres. For $n$ fermions
this is $U(2^k),\,k=\half (n-1),\, n$ odd, or $U(2^k)\xz
U(2^k),\,k=\half(n-2),\, n$ even. It is possible to have gauge group
$U(2^{\half n})$ if one allows all powers of $\h$ in superfield expansions
\cite{Kraus:2000nj}.

Under horizontal diffeomorphisms $A_M$ transforms as

\be
\d A_M=v^N F_{NM}\ .
\la{3.22}
\ee

If one includes the one-form gauge parameter in this calculation one can
also find the corresponding transformation of $A$; it is

\be
\d A_M= D_M z^{\unM} b_{\unM}:=\left(\del_M z^{\unM}+ A_M{}^{\ghm}
\del_{\ghm}z^{\unM}\right)b_{\unM}
\la{3.23}
\ee

In this last equation we have introduced a second covariant derivative

\be
D_M:=\del_M+A_M{}^{\ghm}\del_{\ghm}\ ,\qquad
A_M{}^{\ghm}\del_{\ghm}:=
\d^{\ghm\ghn}A_{M \ghm}\del_{\ghn}\ ,
\la{3.23.1}
\ee

which defines the horizontal
subspaces  in $T\hM$ associated with the Yang-Mills connection.

It is now straightforward to relate $\cF$ to the Yang-Mills field strength $F$
in the standard gauge. One finds

\begin{equation}
\cF_{MN}= F_{MN} -\tB_{MN} -\tB_{M\ghm} N^{\ghm\ghn} \tB_{\ghn N}\ ,
\la{3.24}
\end{equation}

where the tilde denotes a horizontal pull-back constructed using $D_M$ in place
of $\cD_M$. Note that in the standard gauge \eq{3.16} $N$ is the inverse of

\be
K_{\ghm\ghn}=\d_{\ghm\ghn}-B_{\ghm\ghn}
\la{3.25}
\ee

Equation \eq{3.24} gives the generalisation of the modified field strength of a
single D-brane in the non-abelian case, at least in this classical
approximation.


\section{D9-brane}


In this section we shall discuss the D9-brane in  a flat IIB background.
We shall go directly to the physical gauge and compute the Yang-Mills field
strength in the standard flat $N=1, D=10$ superspace basis. The computation
is similar to the abelian case \cite{Akulov:1998bq,Kerstan:2002au}.
The only non-zero component of the torsion is

\be
T_{\ua\ub}{}^{\unc}\rightarrow  T_{\a i\b j}{}^c=-i \d_{ij} (\c^c)_{\a\b}\ ,
\la{5.1}
\ee

where the notation indicates that the underlined $32$-component odd index is
replaced by a doublet of $16$-component Majorana-Weyl indices. Furthermore,
since there are no even directions transverse to the brane we can simply
identify the even tangent spaces of $M$ and $\unM$. In the physical gauge we
shall identify $\th^{\a 1}$ with the odd coordinate of the brane.  The NS
three-form $H$ is taken to be

\bea
H &=& \half E^c E^{\ub} E^{\ua} H_{\ua\ub c}\nn \\
&=& -\frac{i}{2} E^c E^{\b j} E^{\a i} (\c_c)_{\a\b} (\t_1)_{ij}\nn\\
&=&  -i E^c E^{\b 2} E^{\a 1} (\c_c)_{\a\b}\ ,
\la{5.2}
\eea

where $\t_1$ is a Pauli matrix and where

\bea
E^{\a i}&=& d\th^{\a i}\ ,\nn\w1
E^a&=& d x^a-\frac{i}{2} d\th^{\a i}(\c^a)_{\a\b}\th^{\b i}
\la{5.2.1}
\eea

are the usual supersymmetric basis one-forms of flat IIB superspace. A
permissible choice for the potential two-form $B$ is

\bea
B_{a \b 1}&=& i (\c_a \th^{2})_{\b}\nn\\
B_{\a 1\b 2}&=& \frac{1}{3} (\c^a \th^{2})_{\a}(\c_a \th^{2})_{\b}\ ,
\la{5.3}
\eea

with all other components being zero. Here $(\c^a \th^{2})_{\a}=
(\c^a)_{\a\b}\th^{\b 2}$. An advantage of this choice is that this $B$ is
manifestly invariant under the first supersymmetry since it does not depend on
$x$ or $\th^1$. This means that we shall not have to include any specific 
boundary terms in order to ensure that this symmetry holds.

We now come to the gauge-fixing. We shall choose the standard gauge for $\cA$,
i.e. $\cA_{\ghm}=\frac{1}{2} \h_{\ghm}$, and we shall also choose the physical
gauge for the string sigma model field on the boundary, $z^{\unM}(z,\h)$. Thus
we set

\be
\unx^m = x^m \qquad {\rm and }\qquad \unth^{\a 1}=\th^{\a}\ ,
\la{5.4}
\ee

while

\be
\unth^{\a 2}=\L^{\a}(x,\th,\h)\ .
\la{5.5}
\ee

In these two equations we have underlined the target space coordinates for
clarity. The spinor field $\L$ contains all of the covariant physical fields.
Its leading component in the $\h$-expansion corresponds to the $U(1)$ gauge
multiplet which describes the motion of the centre of mass. The higher terms
will correspond to the other, non-abelian, gauge multiplets.

We note that the pull-back of $B$ in the purely vertical direction vanishes in
this gauge. We have

\be
B_{\ghm\ghn}=(\del_{\ghn}z)^{\unB} (\del_{\ghm}z)^{\unA}  B_{\unA\unB}\ .
\la{5.5.1}
\ee

However, only $\th^2$ depends on $\h$, and so $B_{\ghm\ghn}$ must vanish as
$B_{a \b2}=B_{\a 2\b 2}=0$. This in turn implies that

\be
K_{\ghm\ghn}=\d_{\ghm\ghn}\qquad \Rightarrow \qquad  N^{\ghm\ghn}=
\d^{\ghm\ghn}\ . \la{5.5.2}
\ee

We can therefore raise or lower $\ghm$ indices using $\d$ without fear
of ambiguity. We shall also need the covariant pull-backs of $B$ with respect to
$D_M$ in order to compute $F_{MN}$ from \eq{3.24}. We can compute these directly
in the flat basis $e_A=e_A{}^M\del_M=(\del_a,d_{\a})$ where  we have used the
notation $d_{\a}$ for the usual flat superspace covariant
derivative in order to avoid confusion with the Yang-Mills derivative $D_M$. We
shall denote the non-vanishing pull-backs of $B$ in this basis by $\tb_{AB}$ and
$\tb_{A\ghm}$. Thus

\bea
\tb_{AB}&=&(D_B z)^{\unB} (D_A z)^{\unA} B_{\unA\unB}\ ,\nn \w1
\tb_{A\ghm}&=&(\del_{\ghm} z)^{\unB} (D_A z)^{\unA} B_{\unA\unB}\ ,
\la{5.5.3}
\eea

where we emphasise that $D_A$ includes the Yang-Mills field, $D_A=e_A{}^M D_M$.
A straightforward calculation yields

\bea
\tb_{\a\b}&=& \frac{1}{3}D_{(\a}\L\c^a\L (\c_a\L)_{\b)}\ ,\nn\w1
\tb_{a\b}&=&i(\c_a\L)_{\b} + \frac{1}{6} D_a \L \c^b \L
(\c_b \L)_{\b}\,\nn\w1
\tb_{ab} &=&    0\ .
\la{5.5.4}
\eea

One also finds

\bea
\tb_{\a\ghm}&=& \frac{1}{6} \del_{\ghm}\L\c^a\L(\c_a \L)_{\a}\ ,\nn\w1
\tb_{a\ghm}&=& 0 \ .
\la{5.5.5}
\eea

Our strategy now is to evaluate the non-abelian field strength tensor
$F_{MN}$ from \eq{3.24} using the constraints \eq{2.17} and \eq{2.18}. These
constraints are written with respect to a preferred basis in $M$ which is
determined by an induced supervielbein, whereas we want to compute the
components of $F$ with respect to the flat basis in $N=1$ superspace. We can
always choose a basis $E_{A}$ of vector fields in $M$ to have the following
form,

\bea
E_{\a}&=& d_{\a} + \psi_{\a}{}^{b} \del_b \ , \nn \w1
E_a&=& (A^{-1})_a{}^b \del_b \ .
\la{5.6}
\eea

The dual one-form relations are

\bea
E^{\a}&=&e^{\a}\ ,\nn\w1
E^{a}&=&(e^{b}-e^{\b}\psi_{\b}{}^b)A_b{}^a\ ,
\la{5.6.1}
\eea

where $e^{\a}=d\th^{\a},e^a=dx^a-\frac{i}{2} d\th \c^a \th$. The
generalised superembedding matrix \eq{2.19} can be written

\be
\cE_A{}^{\unA}=E_A{}^M \left(D_M + B_M{}^{\ghm}\del_{\ghm}\right)z^{\unM}
E_{\unM}{}^{\unA}\ .
\la{5.7}
\ee

Using \eq{5.6} in \eq{5.7} we find

\be
\cE_{\a}{}^{\una}= -\frac{i}{2}(D_{\a}\L \c^a\L) +
\psi_{\a}{}^b(\d_b{}^a-\frac{i}{2}D_b\L\c^a\L)
-\frac{i}{2}E_{\a}{}^M B_M{}^{\ghm}\,\del_{\ghm}\L\c^a\L\ .
\la{5.8}
\ee

The last term can be written

\bea
-\frac{i}{2}E_{\a}{}^M B_M{}^{\ghm}\del_{\ghm}\L\c^a\L &=& -\frac{i}{2}
\tb_{\a}{}^{\ghm}\,\del_{\ghm}\L\c^a\L\nn\w1
&=&-\frac{i}{12}\chi_{\ghm}{}^a  \chi^{\ghm b} (\c_b\L)_{\a}\ ,
\la{5.9}
\eea

where we have introduced the abbreviation

\be
\chi_{\ghm}{}^a:=\del_{\ghm}\L\c^a\L\ .
\la{5.10}
\ee

The constraint $\cE_{\a}{}^{\una}=0$ therefore allows us to solve for $\psi$,

\be
\psi_{\a}{}^a=\frac{i}{2}(D_{\a}\L\c^b\L +
\frac{1}{6}\chi^{\ghm b}\chi_{\ghm}{}^c
(\c_c\L)_{\a})(\d_b{}^a-\frac{i}{2}D_b\L\c^a\L)^{-1}\ .
\label{5.11}
\ee

To find $A$ we can choose

\be
\cE_a{}^{\una}=\d_a{}^{\una}
\label{5.12}
\ee

since there are no even transverse directions. From this one immediately finds

\be
A_a{}^b= \d_a{}^b-\frac{i}{2}D_a\L\c^b\L\ .
\label{5.13}
\ee

We now turn to the $\cF$ constraint. From equation \eq{3.24} we have

\be
f_{AB}=e_B{}^N e_A{}^M \cF_{MN} +\tb_{AB}+\tb_{A}{}^{\ghm}\tb_{\ghm B}\ ,
\la{5.14}
\ee

where $f_{AB}$ denotes the components of the Yang-Mills field strength in the
flat $N=1$ basis. The first term on the right can be evaluated
straightforwardly using \eq{2.18} and \eq{5.6.1},

\bea
\cF&=&\half E^b E^a \cF_{ab}\nn\w1
&=&\half e^{\b}e^{\a}\psi_{\a}{}^a  \psi_{\b}{}^b A_a{}^c A_b{}^d \cF_{cd}\nn\w1
&&-e^b e^{\a} \psi_{\a}{}^a A_a{}^c A_b{}^d \cF_{cd}+\half
e^{b} e^{a} A_a{}^c A_b{}^d \cF_{cd}\ .
\la{5.15}
\eea

Using the fact that $b_{ab}=b_{a\ghm}=0$ we find

\be
f_{ab}= A_a{}^c A_b{}^d \cF_{cd}\ .
\la{5.16}
\ee

With the aid of \eq{5.5.4} we thus have

\bea
f_{\a\b}&=&\frac{1}{3} D_{(\a}\L\c^a\L (\c_a\L)_{\b)} +\psi_{\a}{}^a
\psi_{\b}{}^b f_{ab} -\frac{1}{36}\chi_{\ghm}{}^a  \chi^{\ghm b}
(\c_a\L)_{\a}(\c_b\L)_{\b}\ ,\la{5.17} \w1
f_{a\b}&=&i  (\c_a\L)_{\b} +\frac{1}{6}D_a\L\c^b\L  (\c_b\L)_{\b} -\psi_{\b}{}^b
f_{ab}\ .
\la{5.18}
\eea

These two equations are the main results for this section.  In order to bring
them to a slightly more familiar form it is useful to introduce a field
$h_{\a}{}^{\b}$ which is the counterpart of the $h$ field which arises in
superembeddings for single branes. As in the abelian case we can put

\be
\cE_{\a}{}^{\ub}=\cases{ \cE_{\a}{}^{\b 1} &=\,  $\d_{\a}{}^{\b}$ \cr
& \cr
\cE_{\a}{}^{\b 2} &=\, $h_{\a}{}^{\b}$ \cr}
\la{5.19}
\ee

Using \eq{5.6} and \eq{5.7}  we find

\be
h_{\a}{}^{\b}=D_{\a}\L^{\b} + \psi_{\a}{}^b D_b \L^{\b}+\frac{1}{6}
\chi^{\ghm a} \del_{\ghm}\L^{\b}(\c_a\L)_{\a}\ .
\la{5.20}
\ee

With the aid of this formula one can show that

\be
h_{\a}{}^{\b}(\c^a\L)_{\b} =-2i\psi_{\a}{}^a
\la{5.21}
\ee

Equation \eq{5.20} can be inverted to give

\be
D_{\a}\L^{\b}=h_{\a}{}^{\c}\left( \d_{\c}{}^{\b} -
\frac{i}{2}(\c^a\L)_{\c} D_a\L^{\b}\right)
-\frac{1}{6}\del_{\ghm}\L^{\c} \del^{\ghm}\L^{\b} (\c^a\L)_{\c}(\c_a \L)_{\a}
\ .
\la{5.22}
\ee

Using \eq{5.21} in \eq{5.17}, \eq{5.18} we find

\bea
f_{\a\b}&=&\frac{1}{3} D_{(\a}\L\c^a\L (\c_a\L)_{\b)} -\frac{1}{4}h_{\a}{}^{\c}
h_{\b}{}^{\d} (\c^a\L)_{\c}(\c^b\L)_{\d} f_{ab} -\frac{1}{36}\chi_{\ghm}{}^a
\chi^{\ghm b} (\c_a\L)_{\a}(\c_b\L)_{\b}\ ,
\la{5.23}
\w1
f_{a\b}&=&i  (\c_a\L)_{\b} +\frac{1}{6}D_a\L\c^b\L  (\c_b\L)_{\b} -\frac{i}{2}
h_{\b}{}^{\c} (\c^b\L)_{\c} f_{ab}\ .
\la{5.24}
\eea

It is easy to check that our results reduce to those of Kerstan's in the abelian
case \cite{Kerstan:2002au}.
The formulae for $f_{\a\b}$ and $f_{a\b}$ are equations $(33)$ and $(34)$
of his paper (there is a factor of $\frac{i}{2}$ missing from the third term in
(34)). Berkovits and Pershin \cite{Berkovits:2002ag} also agree with Kerstan in
this case.\footnote{The BP result for the D9-brane (up to $\h^2$) has also been
derived in a GS calculation starting from the flat target space action 
\cite{hw}.} One can transform to their
conventions from ours by identifying $\L$ with $-W$, by replacing $\c_a$ by
$-i\c_a$ and by adding a minus sign for each commutator. To compare with their
results at order $\h^2$ one has to expand our fields to first non-trivial order,
so that $\L\sim -W - \half\h^2 \hW$. If one looks at their equations for
$f_{\a\b}$ and $f_{a\b}$, their equations (4.6) and (4.7), one sees exactly the
same structure as one does by expanding our equations \eq{5.23} and \eq{5.24}
out keeping only the $\h^2$ terms. To do this one has to identify $h$ with the
object BP refer to as $-\frac{1}{4}(\c F)$. Thus our equation \eq{5.22} is
similar to (3.4) in BP and gives their (4.8) when expanded to $\h^2$. The only
terms which are not quite obvious are the commutator terms in BP. However, it is
not difficult to see that these come from the terms in our formalism which have
two contracted $\h$ derivatives, for example, the third term on the right in
\eq{5.23}.  In fact, we even agree on the coefficients of these terms in both
(4.6) (third term on the right in \eq{5.22}) and (4.8) (third term on the right
in \eq{5.23}).



\section{Covariant formulation}



\subsection{Generalised superembeddings}


In section three we have argued that the geometrical equations which
arise from kappa-symmetry should be covariant under the full diffeomorphism
group of $\hM$, although this is not manifest. In this section we shall develop
a manifestly covariant formalism for the geometry of $\hM$ and the map from
$\hM$ to $\unM$ which will allow us to complete the proof of covariance. In
order to do this we shall introduce a structure on $\hM$ which allows us to
distinguish horizontal and vertical directions in a covariant way. We then
introduce a connection for suitable structure group in a similar fashion to the
standard superspace formalism. The resulting geometry is not strictly speaking
that of a gauge bundle over the brane worldvolume. Indeed, the
vertical ($\h$) distribution is not integrable as it would be if we were
reformulating the geometry of a gauge bundle in this way.

We define basis one-forms $\hE^{\hA}$ and dual vector fields $\hE_{\hA}$ in the
usual way by means of a supervielbein and its inverse,

\begin{equation}
\hE^{\hA}:=dz^{\hM} E_{\hM}{}^{\hA}\qquad \hE_{\hA}:= E_{\hA}{}^{\hM}\del_{\hM}
\ .
\label{4.1}
\end{equation}

The structure group is taken to be a product which splits the tangent space into
horizontal and vertical components, so that the splitting
$\hE^{\hA}=(\hE^A,\hE^{\gha})$ is invariant. The horizontal part of the
structure group, acting on $\hE^A$, will be the same group as the superspace
structure group, while the vertical part will turn out to be $O(n)$ and
the parameters of these groups should depend on all of the
coordinates. We then introduce a set of connection one-forms $\O_{\hA}{}^{\hB}$
to allow us to carry out covariant differentiation with respect to the structure
group. The torsion and curvature forms are defined in the usual way,

\bea
T^{\hA}:&=& d \hE^{\hA} + \hE^{\hB} \O_{\hB}{}^{\hA} \nn\w2
R_{\hA}{}^{\hB} :&=& d \O_{\hA}{}^{\hB}  +\O_{\hA}{}^{\hC}\O_{\hC}{}^{\hB}
\la{4.2}
\eea

At this stage we have introduced a lot of new objects and, at the same time, we
have enlarged the gauge group, so that it will be necessary to impose some
constraints. In the brane context we shall be able to derive many of these from
what one might call the generalised superembedding formalism, i.e the geometry
of the map $\hf:\hM\rightarrow \unM$.

To summarise, we now have a superspace $\hM$, equipped with the above structure,
and we also have the two-form $K$ satisfying the Bianchi identity $dK=-\hf^* H$.
We shall impose the following constraints on $K$,

\bea
K_{\gha\ghb}&=&\d_{\gha\ghb}\nn\w2
K_{A\ghb} &=& 0
\la{4.3}
\eea

The first of these is natural in the sense that we do not wish to introduce any
new fields and because this component of $K$ must be non-singular. It also
reflects the odd-symplectic structure of the vertical direction. We shall
further suppose that $\d_{\gha\ghb}$ is covariantly constant so that the
vertical structure group is indeed $O(n)$. The second constraint in \eq{4.3}
corresponds to the fact that $K$ has no mixed components in the horizontal lift
basis.  The remaining  non-trivial part of $K$ is related to the field
$\cF_{AB}$ as we shall discuss later. The $K$ Bianchi identity has the following
components

\bea
3\nab_{[A} K_{BC]} + 3T_{[AB}{}^D K_{|D|C]} &=& - H_{ABC}\ ,\la{4.4}\w2
\nab_{\gha} K_{BC}+ 2 T_{\gha[B}{}^D K_{|D|C]} + T_{BC}{}^{\ghd}K_{\ghd\gha}&=&
-H_{\gha BC}\ ,\la{4.5}\w2
T_{\gha\ghb}{}^{D} K_{DC} + 2 T_{C(\gha}{}^{\ghd} K_{\ghb)\ghd}&=&-H_{\gha\ghb
C}\ ,\la{4.6}\w2
3T_{(\gha\ghb}{}^{\ghd} K_{\ghc)\ghd}&=&-H_{\gha\ghb\ghc}\ .
\la{4.7}
\eea

The covariant derivative here is constructed using the connection we have just
introduced and the terms on the right-hand-sides are the various components of
the pull-back of $H$ in this basis. The first of these equations is equivalent
to  \eq{3.7}, while the second can be rearranged to give

\begin{equation}
T_{AB}{}^{\ghc}=-N^{\ghc\ghd}\left(H_{AB\ghd}+ \nab_{\ghd}K_{AB}+
2T_{\ghd[A}{}^C K_{|C|B]}\right)\ ,
\la{4.8}
\end{equation}

where we have denoted the inverse of $K_{\gha\ghb}$ by $N^{\gha\ghb}$, although
it is just $\d^{\gha\ghb}$ in this basis. The left-hand-side of \eq{4.8} is
essentially the curvature $\cR$ of $\cD$ so that the equation is the
counterpart of \eq{3.8}. Equation \eq{4.6} allows us to solve for
$T_{A(\ghb\ghc)}$ while \eq{4.7} allows us to solve for $T_{(\gha\ghb\ghc)}$,
(where we have lowered the upper torsion indices using $K_{\gha\ghb}$). The
remaining parts of these components of the torsion may be set to zero using the
freedom to choose $\O_{A\ghb\ghc}$ and  $\O_{\gha\ghb\ghc}$, both of which are
antisymmetric on the last two (Lie algebra) indices. It will turn out that all
of the remaining components of the torsion which are so far not determined in
terms of known quantities can be found from the torsion equation of the
generalised superembedding.

One can continue with an analysis of the curvature tensor but we shall not do so
in any detail here as it will not be needed in the rest of the paper. However,
it is worth making one point which arises because the structure group in the
fermionic direction is orthogonal which is symplectic for odd variables. In 
this case one finds that the torsion constraints do not determine the purely
fermionic connection $\O_{\gha,\ghb}{}^{\ghc}$ completely. The totally
antisymmetric part is left over. This should be regarded as being pure gauge,
i.e. we should introduce a shift symmetry precisely of this form. If we then set
the curvature $R_{\gha\ghb,\ghc}{}^{\ghd}$ to zero, and use the $O(n)$ gauge
symmetry to set the connection to zero, the additional shift symmetry will allow
us to preserve the gauge choice without overconstraining the residual gauge
parameters.

We now
give a brief discussion of the torsion equation for the generalised
superembedding. The embedding matrix can be defined, as in the case of a single
brane, to be the derivative of the map $\hf$ in the preferred bases for both the
target space and $\hM$. Thus we have

\be
\hE_{\hA}{}^{\unA}=  \hE_{\hA}{}^{\hM}\del_{\hM}z^{\unM} E_{\unM}{}^{\unA}\ .
\la{4.9}
\ee

The torsion equation is

\be
2\nab_{[\hA}\hE_{\hB]}{}^{\unC} + T_{\hA\hB}{}^{\hC} \hE_{\hC}{}^{\unC}=
\hE_{\hB}{}^{\unB}\hE_{\hA}{}^{\unA}T_{\unA\unB}{}^{\unC}\ .
\la{4.10}
\ee

We can impose some constraints on the superembedding matrix. The first is the
constraint \eq{2.17}. The remaining parts of $\hE_{A}{}^{\unA}$ can be
parametrised in a similar fashion to the single brane case. Thus we can take

\bea
\hE_{\a}{}^{\una} &=& 0\ , \nn \w1
\hE_{a}{}^{\una}  &=& u_a{}^{\una}\ ,\nn\w1
\hE_{\a}{}^{\ua} &=& u_{\a}{}^{\ua} + h_{\a}{}^{\b'} u_{\b'}{}^{\ua}\ ,\nn\w1
\hE_{a}{}^{\ua}&=& h_{a}{}^{\b'} u_{\b'}{}^{\ua}\ ,
\la{4.11}
\eea

where $u$ denotes an element of the target space structure group in the
appropriate representation. We shall also impose some constraints on the new
components of the superembedding matrix. They are

\bea
\hE_{\gha}{}^{\una} &=& h_{\gha}{}^{a'} u_{a'}{}^{\una}\ , \nn\w1
\hE_{\gha}{}^{\ua} &=& h_{\gha}{}^{\a'} u_{\a'}{}^{\ua}\ .
\la{4.12}
\eea

In \eq{4.12} the $\gha$ index can be thought of as a derivative in the fermionic
direction, so that these equations imply that the superembedding only depends on
the $\h$-coordinates in the transverse directions. This is
necessary in order to get the right number of degrees of freedom on the brane.
In the discussion of section three it was stated that the latter was a result of
the $\h$-dependence of the diffeomorphisms of $M$, but the situation is slightly
different in this version as we have introduced new components for the
supervielbein which ``use up'' some of the gauge transformations so that new
constraints have to be imposed. The fields $h_{\a}{}^{\a'}$ and $h_{a}{}^{\a'}$
can be thought of as $M$ derivatives of the transverse fermion field. The first
is related to the field-strength $\cF_{ab}$ (or $K_{ab}$) by the $\cF$ Bianchi
identity.

As in the standard superembedding formalism  the connection $\O_{\hA,B}{}^C$ can
be found by introducing the Lie algebra-valued one-form $X_{\hA}$,

\begin{equation}
X_{\hA}:= (\nab_{\hA}u) u^{-1}\ ,
\label{4.12.1}
\end{equation}

and setting some parts of it to zero. That is, we put

\begin{equation}
X_{\hA,b}{}^c=X_{\hA,b'}{}^{c'}=X_{\hA,\b}{}^{\c}=X_{\hA,\b'}{}^{\c'}=0\ .
\label{4.13.1}
\end{equation}

These equations determine the connection $\O_{\hA,B}{}^C$. The connection
$\O_{\hA,\ghb}{}^{\ghc}$ is determined by the torsion constraints discussed
above, apart from the totally antisymmetric part of $\O_{\gha,\ghb\ghc}$ which
can be set to zero as explained previously. The torsion equations can then be
solved systematically to determine the remaining components of the
torsion. In addition one finds constraints on the physical fields and their
derivatives. As there are no auxiliary fields and we have manifest supersymmetry
it follows that these equations determine the dynamics of the brane multiplet.


\subsection{Proof of covariance}


The formalism just introduced is manifestly covariant by construction, but it is
not quite so obvious how it is related to the earlier geometrical description
given in section 3. In  the new formalism we have components of
the supervielbein in the boundary fermion directions which were not present
before, and one might wonder how the old formalism could give covariant
results. In particular, the definition of the embedding matrix in the new
formalism differs from the earlier one. In order for them to be the same (for
$\hA=A$) one would have to identify $E_A{}^{\ghm}$ with $-E_A{}^M K_M{}^{\ghm}$,
but this is not the case as we shall now discuss in more detail.

We set

\bea
E_{\gha}{}^M&=& E_{\gha}{}^{\ghm} \f_{\ghm}{}^M \nn \w1
E_{A}{}^{\ghm}&=& E_{A}{}^{M} \psi_{M}{}^{\ghm}\ .
\la{4.13}
\eea

Then the constraint $K_{\gha B}=0$ can be written

\be
K_{\gha B}=E_B{}^{N} E_{\gha}{}^{\ghm}\left(K_{\ghm N} -\psi_N{}^{\ghn}
K_{\ghn\ghm}+ \f_{\ghm}{}^M (K_{MN} +\psi_N{}^{\ghn} K_{M\ghn})\right)=0\ .
\la{4.14}
\ee

This can be solved to give

\be
\psi_M{}^{\ghm}=-K_M{}^{\ghm} + \F^{\ghm N}\cF_{NM}\ ,
\la{4.15}
\ee

where $\cF_{MN}$ is defined in \eq{2.14} and

\be
\F^{\ghm N}:=(\d^{\ghm}{}_{\ghn} +\f^{\ghm N} K_{N \ghn})^{-1}\f^{\ghn N}\ .
\la{4.16}
\ee

A similar calculation for $K_{AB}$ yields

\be
K_{AB}=E_B{}^N E_A{}^M(\cF_{MN} +\cF_{MP}\F^{PQ}\cF_{QN})\ ,
\la{4.17}
\ee

where

\be
\F^{PQ}:=\F^{\ghm P}K_{\ghm\ghn} \F^{\ghn Q}\ .
\la{4.18}
\ee

The horizontal projection of the superembedding matrix can be written as

\bea
\hE_{A}{}^{\unA}&=&E_A{}^{\hM} \del_{\hM} z^{\unM} E_{\unM}{}^{\unA} \nn\w2
&=& E_A{}^M(\del_M + \psi_M{}^{\ghm}\del_{\ghm})z^{\unM} E_{\unM}{}^{\unA}\nn\w2
&=& E_A{}^M(\cD_M + \F^{\ghm N}\cF_{NM}\del_{\ghm})z^{\unM}
E_{\unM}{}^{\unA}\ ,
\la{4.19}
\eea

so that the difference between the old and new definitions resides
in the term with $\cF$. Explicitly,

\be
\hE_{A}{}^{\unA}=\cE_{A}{}^{\unA} +   E_A{}^M
\F^{\ghm N}\cF_{NM}\del_{\ghm}z^{\unM} E_{\unM}{}^{\unA}\ .
\la{4.20.1}
\ee

We also have

\be
K_{\a B}=\cF_{\a C}(\d^C{}_B + \F^{CD} \cF_{DB})\ .
\la{4.20}
\ee

Therefore the original kappa-symmetry constraints

\be
\cE_{\a}{}^{\una}=\cF_{\a B}=0
\la{4.21}
\ee

are precisely
equivalent to the manifestly covariant constraints

\be
\hE_{\a}{}^{\una}=K_{\a B}=0\ .
\la{4.22}
\ee

This discussion makes it clear that the truly covariant  $U(1)$ field strength
is $K$ rather than $\cF$. As we have seen the constraints $K_{\a B}=0$ are
equivalent to the constraints $\cF_{\a B}=0$, but the invariantly
defined non-vanishing components are $K_{ab}$ rather than $\cF_{ab}$.


\section{Discussion}


In this paper we have presented a preliminary study of a possible approach to
finding the non-abelian counterpart to the supersymmetric Born-Infeld plus
Wess-Zumino term action which describes the low energy dynamics of single
D-branes
\cite{Cederwall:1996pv,Aganagic:1996pe,Cederwall:1996ri,
Bergshoeff:1996tu,Aganagic:1996nn}. The
next step in the programme would be to quantise the boundary fermions. If it is
possible to do so consistently, one would then arrive at the desired theory
which would presumably represent a well-defined approximation to the
effective action (or equations of motion) of a set of coincident D-branes. The
full effective action would include higher derivative contributions which in
our approach  should correspond to the quantisation of
the bulk part of the action. Since we do not know how to do this for the GS
action, it would be necessary to use the pure spinor formalism for this purpose.

At the level at which we are working, however, the GS/superembedding approach
has the advantage  that all the symmetries are manifestly realised. In
particular, we have been able to achieve gauge invariance for the $B$-field as
well as kappa-symmetry. As we noted in the introduction, previous attempts in
this direction have not been completely successful, but in our approach
kappa-symmetry with a parameter which depends on the boundary fermions comes out
naturally. In the quantised version this would correspond to a matrix-valued
kappa parameter as advocated in \cite{Bergshoeff:2000kt}. As we  have seen,
the way that kappa-symmetry works is quite complicated;
the use of the equations of motion for the boundary fermions introduces
a vertical/horizontal splitting in $\hM$ as a consequence of which the
covariance of the kappa-symmetry contraints under diffeomorphisms
of $M$ which depend on the boundary fermions is not at all guaranteed. The fact
that everything works out consistently is a strong
indication in favour of the formalism. It would be interesting to investigate if
the pure spinor string with boundary fermions could be analysed in a covariant
fashion with nilpotence of the BRS operator as a replacement for kappa-symmetry.
Since this can be done for type II supergravity \cite{Berkovits:2001ue}
it should also be possible to do it for this case .

In the classical approximation it makes sense to investigate the equations of
motion order by order in the boundary fermions. We have not done this in detail
here, postponing a discussion until the fermions are quantised, but we have seen
that the equations for the non-abelian $N=1,D=10$ superspace field strength for
the D9-brane  in a flat background  (which imply the equations of
motion for the component fields) agree with the results of reference
\cite{Berkovits:2002ag}. In the abelian case formally similar-looking
constraints are known to describe supersymmetric Born-Infeld theory and so these
equations must describe a non-abelian version, although it is not so easy to
verify this explicitly. We also note that it is straightforward to adapt the
D9-brane analysis given here to the case of lower-dimensional branes in flat
backgrounds. One of course sees the emergence of non-abelian scalars, here
represented as transverse scalars depending on the boundary fermions. There are
many complicated interaction terms encoded in our formalism and we would
therefore expect to see, for example, the interactions responsible for
the dielectric brane phenomenon appearing at some point.

Another way of investigating such terms would be via an action. GS actions for
D-branes can be constructed by a systematic
procedure in the superembedding formalism starting from a closed $(p+2)$-form on
the superworldvolume \cite{Bandos:1995dw,Howe:1998ts}. This form is directly
related to the WZ term but also encodes the kinetic term via supersymmetry.  In
the approximation we are using in the current paper it is not clear that it
makes sense to write an action. However, it should be possible to investigate
whether there are candidate $(p+2)$-forms which could yield action forms after
quantisation. For a set of D0-branes we need a closed two-form on $M$ in order
to implement the brane action principle. This suggests that we should be looking
for a covariantly closed horizontal two-form, $\cW$ say, which satisfies an
equation of the form

\begin{equation}
{(D\cW)}_{ABC}=0\ ,
\label{6.1}
\end{equation}

where $D$ is a suitably defined covariant exterior derivative which should not
involve torsion terms with any $\gha$-type indices.
To get the action form itself one should quantise and then take the trace over
the gauge indices in $\cW$ in which case we should obtain a closed two-form on
$M$ as required. We expect that $\cW$ should be constructed from the IIA RR
field strengths $G_p,\,p=2,4,\ldots,$ which satisfy $d G_{p+2}=H \wedge G_{p}$
($G_0=0$), and from $K_{AB}$. For the D0-brane things are slightly simpler since
$K_{AB}=H_{ABC}=0$. If we start with $G_2$ we find

\begin{equation}
\nab_{[A} G_{BC]}+ T_{[AB}{}^D G_{|D|C]}=-T_{[AB}{}^{\ghd}G_{\ghd C]}
\label{6.2}
\end{equation}

The term on the right gets in the way, but can be manipulated if we use $dK=0$
to replace $T_{AB}{}^{\hat\d}$ by $-N^{\hat\d\hat\e} H_{AB\hat\e}$. We then see
that the right-hand side resembles $N^{\hat\d\hat\e}(G_2\wedge
H)_{ABC\hat\d\hat\e}$. But $G_2\wedge H=dG_4$, and so we can deal with such a
term by appropriately amending $G_2$. In this way we are naturally led to
consider the following two-form as a candidate $\cW$,

\begin{eqnarray}
{\cW}_{AB}&=&G_{AB}-\half N^{\gha\ghb} G_{\gha\ghb AB}+\frac{1}{8}
N^{\gha\ghb} N^{\ghc\ghd} G_{\gha\ghb\ghc\ghd AB}-\ldots\nn\w1
&=& \left( \exp(-\half i_N) \left(\sum G\right)\right)_{AB}\ ,
\label{6.3}\end{eqnarray}

where $i_N$ denotes contraction with the bivector $N^{\hat\a\hat\b}$. This
expression is exactly what one would expect from the Wess-Zumino term in the
bosonic non-abelian D-brane action since two $\gha$-type indices on a
pulled-back RR form  contracted with $N$ correspond to the $\i_{\phi^2}$
operation of \cite{Myers:1999ps}, although in our case
there will be fermion contributions as well. However, with this definition of
$\cW$ we do not quite get an equation of the form of \eq{6.1} since there are
additional terms left over. These seem to give rise to a divergence in the
fermionic direction which one might expect should vanish in the action since
gauge invariance corresponds to $\h$-independence at the classical level. We
postpone a full discussion of this point until we have quantised the theory, but
it is encouraging that the modifed Wess-Zumino term does seem to emerge from
supersymmetry.  It would also be interesting to see if kappa-symmetry implies
any modifications to the interactions found in
\cite{Myers:1999ps,Taylor:1999gq,Taylor:1999pr} such as those discussed in references
\cite{Hassan:2000zk,Hassan:2003uq}.

We conclude with a few comments on quantisation.  As noted in the introduction,
the dependence of the bulk fields on the boundary fermions means that it is not
straightforward to integrate out the latter from the path integral. One
possibility would be to find an effective boundary lagrangian that would
reproduce the desired boundary equations of motion, but it is not clear that
such an object exists. Another possibility would be to try deformation
quantisation and demand consistency with the symmetries of the problem. There is
a natural bracket in the bounday fermion direction  given by $N^{\gha\ghb}$
which has the advantage of being covariant. However, it is not Poisson which
suggests that it might be appropriate to consider a non-associative star
product. This has been advocated in the bosonic sector in the presence of a
non-trivial spacetime $B$-field \cite{Cornalba:2001sm},
although here it might be necessary even in a
flat background since the superspace $B$-field is never trivial.


\section*{Acknowledgements}

We thank Fawad Hassan and Per Sundell for extensive discussions.
This work was
supported in part by EU grant (Superstring theory) MRTN-2004-512194, PPARC grant
number \linebreak PPA/G/O/2002/00475 and VR grant 621-2003-3454. PSH thanks the 
Wenner-Gren foundation.

\end{document}